\documentclass[3p,times,procedia]{elsarticle}
\usepackage{nupha_ecrc}

\volume{00}

\firstpage{1}

\journalname{Nuclear Physics A}

\runauth{}

\jid{nupha}

\jnltitlelogo{Nuclear Physics A}

\usepackage{amssymb}

\usepackage[figuresright]{rotating}

\begin{document}

\begin{frontmatter}



\dochead{XXVIth International Conference on Ultrarelativistic Nucleus-Nucleus Collisions\\ (Quark Matter 2017)}

\title{A Unified Description for Comprehensive Sets of \\
Jet Energy Loss Observables with CUJET3}


\author[IU]{Shuzhe Shi}
\author[IU,CU]{Jiechen Xu}
\author[IU,CCNU]{Jinfeng Liao}
\author[LBL,CU,CCNU]{Miklos Gyulassy}

\address[IU]{Physics Department and Center for Exploration of Energy and Matter,
Indiana University, \\ 2401 N Milo B. Sampson Lane, Bloomington, IN 47408, USA}

\address[LBL]{Nuclear Science Division, Lawrence Berkeley National Laboratory, Berkeley, CA 94720, USA}

\address[CU]{Pupin Lab MS-5202, Department of Physics, Columbia University, New York, NY 10027, USA}

\address[CCNU]{Institute of Particle Physics, Central China Normal University, Wuhan, China}

\begin{abstract}
Jet energy loss in heavy ion collisions, as quantified by the traditional observable of high $p_T$ hadron's nuclear modification factor $R_{AA}$, provides highly informative ``imaging'' of the hot medium created in heavy ion collisions. There are now comprehensive sets of available data, from average suppression to azimuthal anisotropy, from light to heavy flavors, from RHIC 200GeV to LHC 2.76TeV as well as 5.02TeV collisions. A unified description of such comprehensive data presents a stringent vetting of any viable model for jet quenching phenomenology. In this contribution we report such a systematic and successful test of CUJET3, a jet energy loss simulation  framework built upon  a nonperturbative microscopic model for the hot medium as a semi-quark-gluon-monopole plasma (sQGMP) which integrates two essential elements of confinement, i.e. the Polyakov-loop suppression of quarks/gluons and emergent magnetic monopoles. 
\end{abstract}


\end{frontmatter}


\section{Introduction}
\label{}

Highly energetic jets, produced by initial hard scatterings in a heavy ion collision, provide invaluable tomographic ``imaging'' of the hot bulk medium created in such collision. A conventional observable for quantifying medium attenuation of jets is the high $p_T$ hadron's nuclear modification factor $R_{AA}$ defined as:
\begin{eqnarray}
R_{AA}\left (p_T,\phi; b; \sqrt{s}; f \right ) = \frac{\frac{dN_{AA}}{p_Tdp_Td\phi}}{T_{AA} \frac{d\sigma_{pp}}{p_Tdp_Td\phi}} \simeq R_{AA}\left (p_T; b; \sqrt{s}; f \right ) \left[ 1 + 2 v_2\left (p_T; b; \sqrt{s}; f \right ) \cos(2\phi)\right]
\end{eqnarray}
where $p_T$ and $\phi$ are transverse momentum and azimuthal angle of observed leading hadrons, with $b$, $\sqrt{s}$ and $f$ specifying the collision centrality, beam energy as well as hadron flavor species e.g. light flavor pions and heavy flavor D mesons. 
A deviation of $R_{AA}$ from unity indicates medium modification in AA collisions that would be absent in pp collisions. Furthermore, due to different in-medium  path length for jets penetrating the medium along different azimuthal directions, a nontrivial $\phi$-dependence shall arise from jet attenuation, specifically with a dominant elliptic component quantified by the coefficient $v_2$. 

Strong suppression effect (with $R_{AA}$ significantly less than one) as well as sizable anisotropy $v_2$ at high transverse momentum have been consistently observed from RHIC to LHC~\cite{Khachatryan:2016odn,ATLAS:2017rmz,Sirunyan:2017pan,Abelev:2012hxa,Abelev:2012di,ATLAS:2011ah,Chatrchyan:2012xq,Adare:2008qa,Adare:2012wg,CMS:2016nrh,CMS:2016jtu,ALICE:2012ab,Abelev:2014ipa}.  There are now comprehensive sets of available data, from average suppression to azimuthal anisotropy, from light to heavy flavors, from RHIC 200GeV to LHC 2.76TeV as well as 5.02TeV collisions. A unified description of such comprehensive data presents a stringent vetting of any viable model for jet quenching phenomenology, as indeed demonstrated by past studies. For example the azimuthal anisotropy was found  to pose severe challenge for models and hint at highly nontrivial temperature dependence of jet-medium coupling~\cite{Liao:2008dk,Zhang:2012ha,Zhang:2012ie,Das:2015ana}. The beam-energy dependence was also found to suggest a considerable reduction of average medium opaqueness from RHIC to LHC (which again hints at strong temperature dependence)~\cite{Horowitz:2011gd,Betz:2012qq,Zhang:2012ha,Zhang:2012ie,Burke:2013yra}. The puzzling close proximity between light flavor and heavy flavor energy loss was found to indicate at the necessity of including elastic energy loss and magnetic screening effect~\cite{Buzzatti:2011vt,Xu:2014ica}. Currently there are a number of jet energy loss modeling frameworks differing in their implementation of hot medium and energy loss scheme with varied degrees of sophistication~\cite{JET,Armesto:2011ht,Chien:2015vja,Bianchi:2017wpt,Cao:2017hhk,Xu:2014tda,Xu:2015bbz,SXLG}, and the large amount of high precision data will be a great opportunity to quantitatively analyze the phenomenological viability of each framework.

In this contribution we report such a systematic and successful test of CUJET3~\cite{Xu:2014tda,Xu:2015bbz,SXLG}, a jet energy loss simulation  framework built upon  a non-perturbative microscopic model for the hot medium as a semi-quark-gluon-monopole plasma (sQGMP) which integrates two essential elements of confinement near the transition temperature $T\to T_c^+$, i.e. the emergent magnetic monopoles~\cite{Liao:2006ry,DAlessandro:2007lae}  and  the Polyakov-loop suppression of quarks/gluons~\cite{Hidaka:2008dr} in the near-$T_c$ plasma.

\begin{figure}[!hbt]
\begin{center} \vspace{-0.1in}
\includegraphics[width=0.9\textwidth]{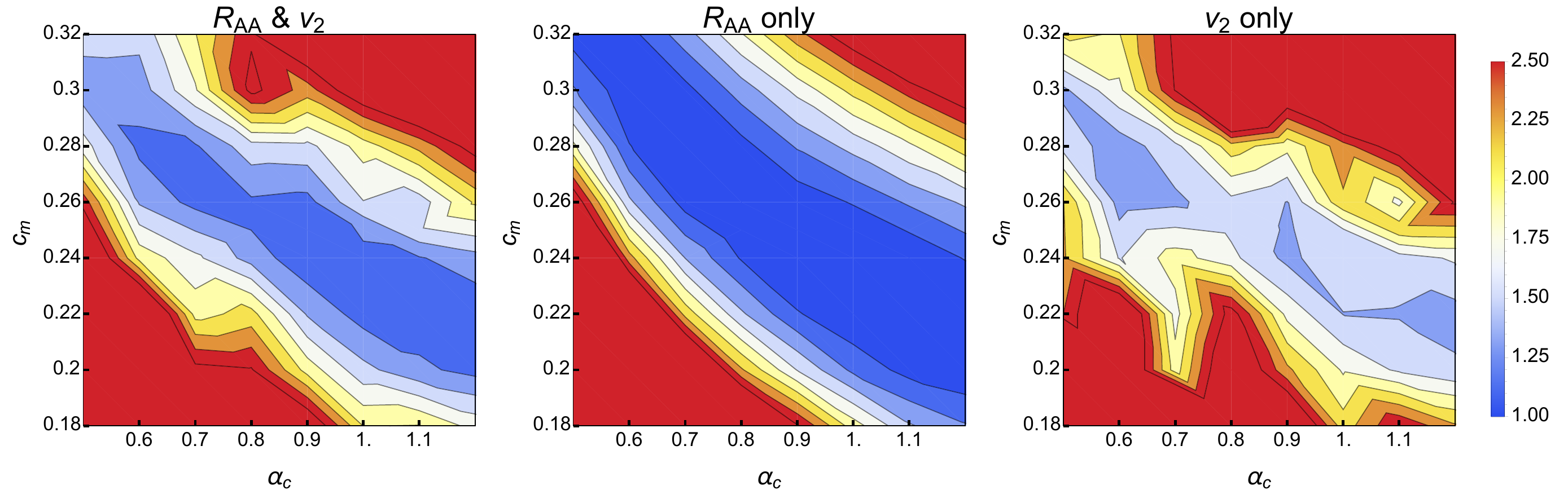} 
\vspace{-0.15in}
\caption{(color online) The $\chi^2/d.o.f$ distribution on $(\alpha_c,c_m)$ parameter plane, from comparing CUJET3 results for pion high $p_T$ observables with central and semi-central data from RHIC 200GeV, LHC 2.76TeV as well as 5.05TeV collisions: (left) including both $R_{AA}$ and $v_2$ data; (middle) including only $R_{aa}$; (right) including only $v_2$.}  \label{fig_chi}
\end{center}
\vspace{-0.4in}
\end{figure} 

\section{Constraining CUJET3 Model Parameters with Light Flavor Data}
\label{}

In the CUJET3 framework, there are two key model parameters. The first is $\alpha_c$ which is the value of QCD running coupling at the non-perturbative scale $T_c$ and sensitively controls the overall opaqueness of the hot medium. The other is $c_m$ which is the coefficient for magnetic screening mass in the medium and influences the contribution of the magnetic component to the jet energy loss. See details of CUJET3 in \cite{Xu:2015bbz}. 

\begin{figure}[!hbt]
\begin{center} \vspace{-0.1in}
\includegraphics[width=0.35\textwidth]{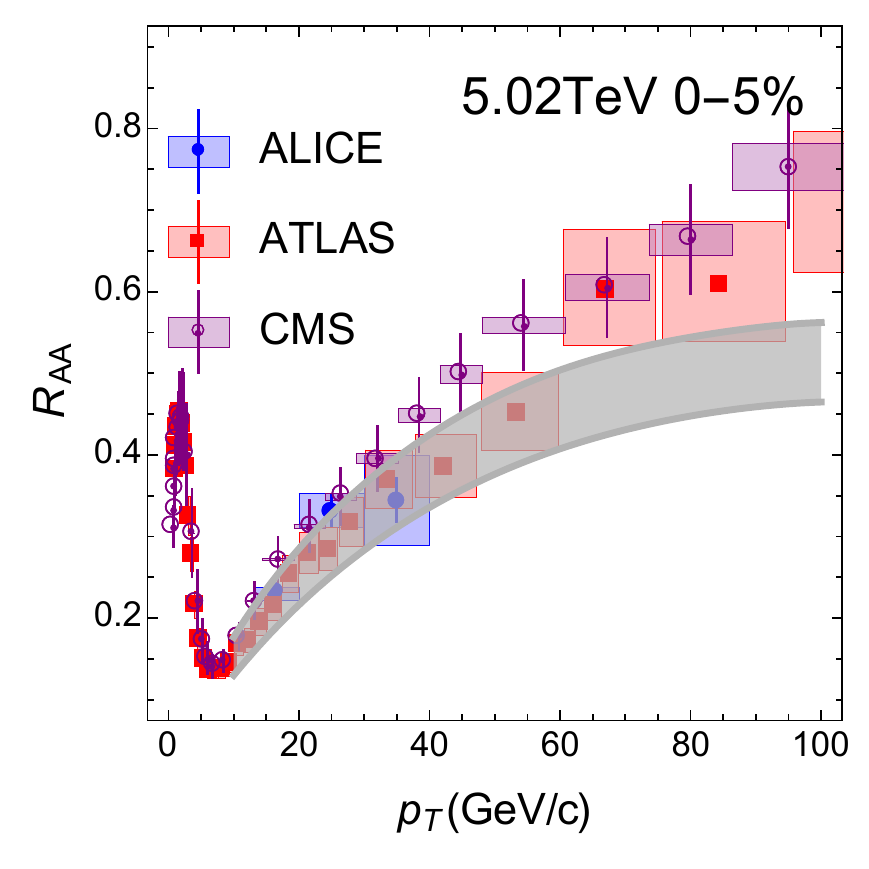} \hspace{0.02in}
\includegraphics[width=0.35\textwidth]{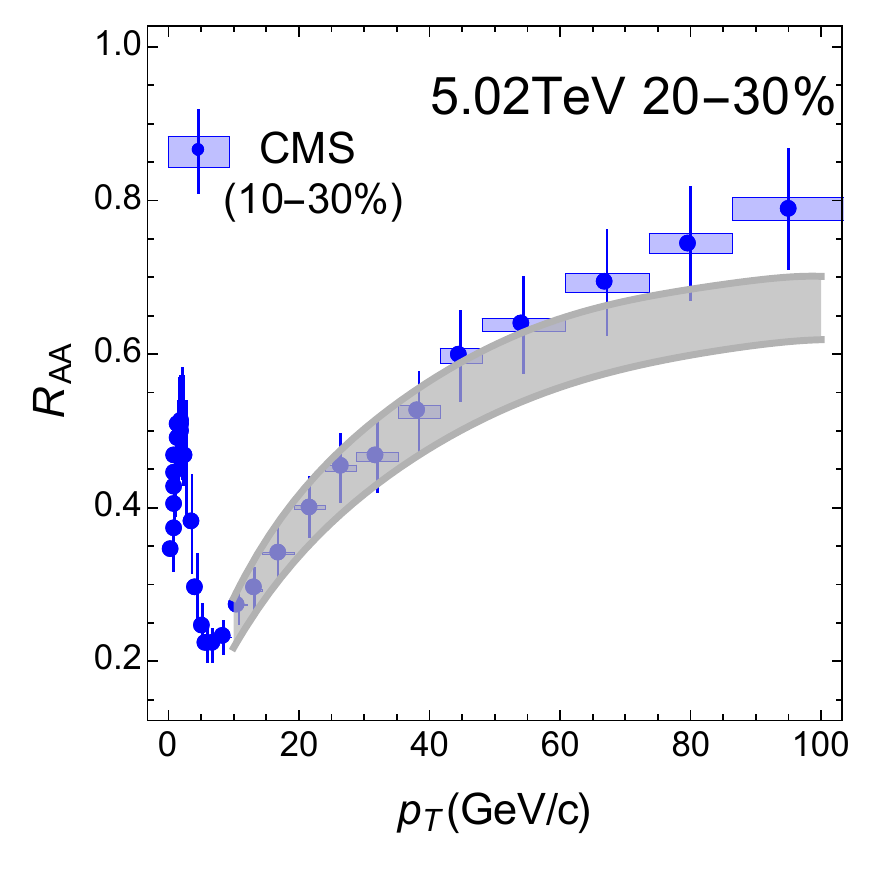} \hspace{0.02in} \\
\includegraphics[width=0.35\textwidth]{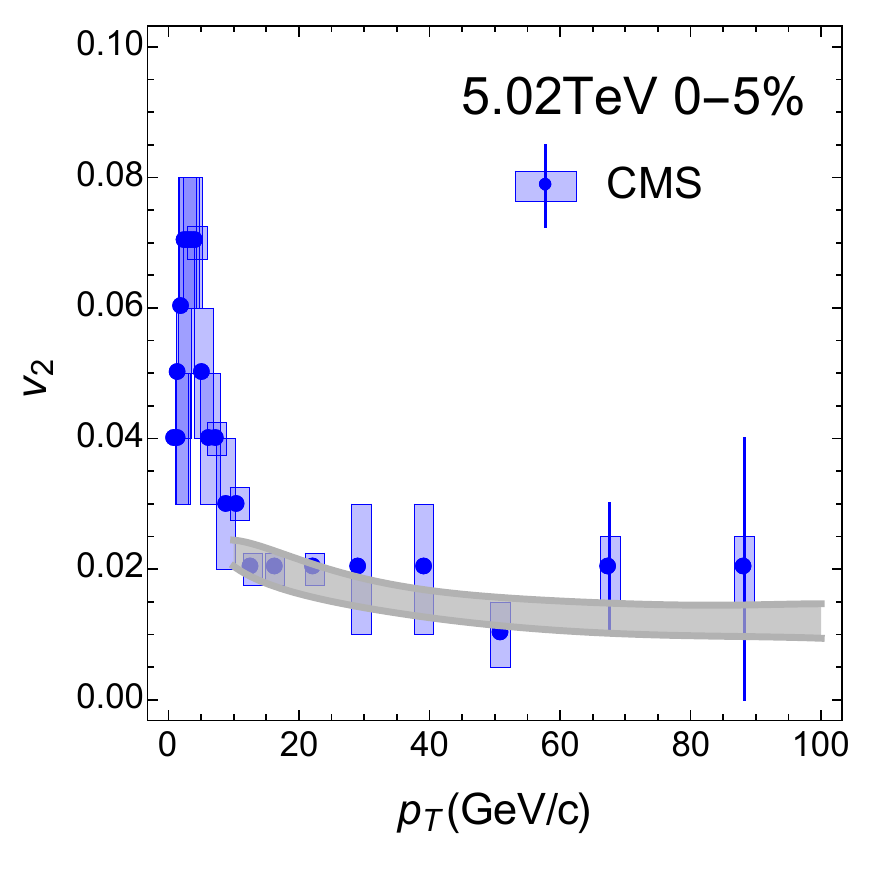} \hspace{0.02in}
\includegraphics[width=0.35\textwidth]{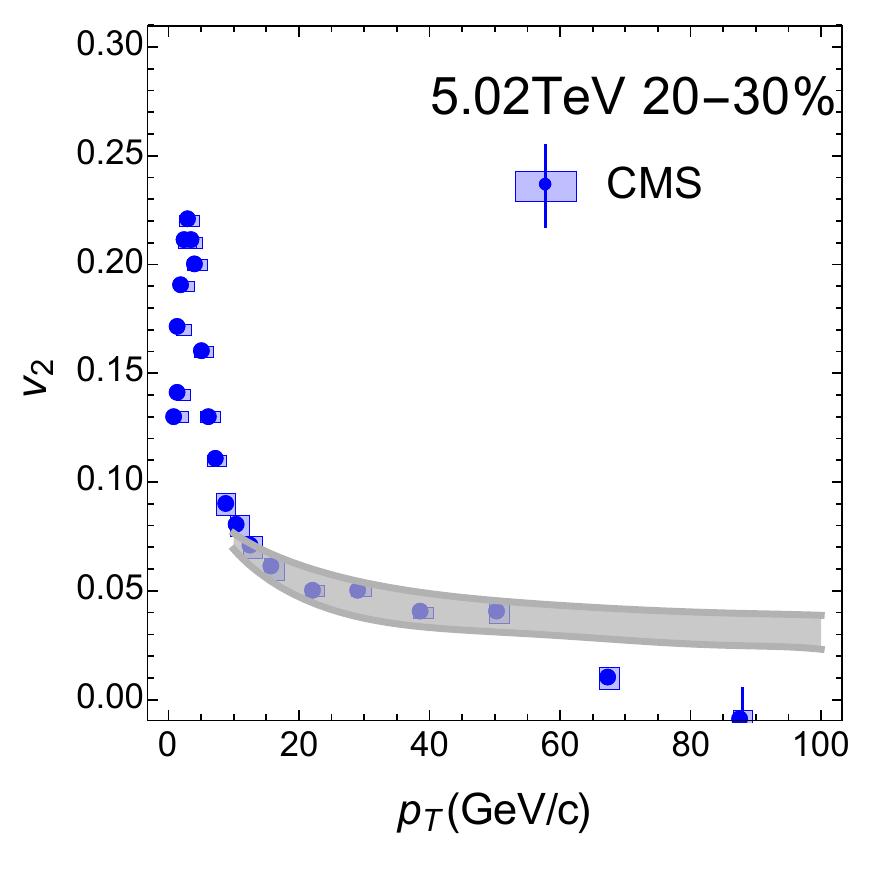}
\vspace{-0.2in}
\caption{(color online) The central and semi-central $R_{AA}$ and $v_2$ of high $p_T$ pions at LHC 5.02TeV collisions, computed from CUJET3 and compared with currently available experimental data.}  \label{fig_light}
\end{center}
\vspace{-0.3in}
\end{figure}

A first step we take is to utilize  central and semi-central high $p_T$ light hadron's $R_{AA}$ and $v_2$ data for all three collision energies to systematically constrain the two key model parameters by a quantitative $\chi^2/d.o.f$ analysis: see Fig.~\ref{fig_chi}. One useful insight is that different observables have different constraining power: the $\chi^2/d.o.f$ distribution for analysis with only $R_{AA}$ data (middle panel) or only $v_2$ data (right panel) favors different regions of parameter space. Taking all data together (left panel), we identify a data-selected optimal parameter region spanned by $(\alpha_c=0.8,c_m=0.22)$ and $(\alpha_c=1.0,c_m=0.28)$ with a $\chi^2/d.o.f$ close to unity. With such constrained parameters, we show in Fig.~\ref{fig_light} the comparison between CUJET3 results with experimental data for  the central and semi-central $R_{AA}$ and $v_2$ of high $p_T$ pions at LHC 5.02TeV collisions. One clearly sees excellent agreement for 5.02TeV data, which demonstrates the newest successful test of CUJET3 in addition to the similar success with 2.76TeV and 200GeV data test already reported in ~\cite{Xu:2014tda,Xu:2015bbz}.

\section{Testing CUJET3 with Heavy Flavor Measurements}
\label{}

\begin{figure}[!hbt]
\begin{center} \vspace{-0.1in}
\includegraphics[width=0.35\textwidth]{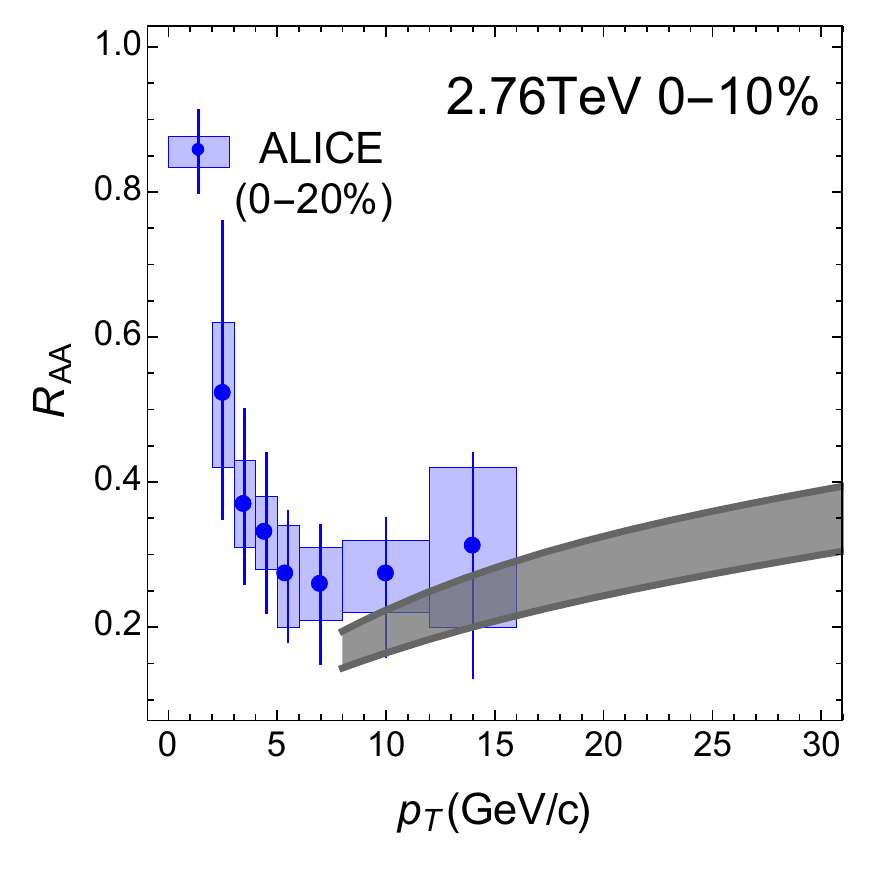} \hspace{0.02in}
\includegraphics[width=0.35\textwidth]{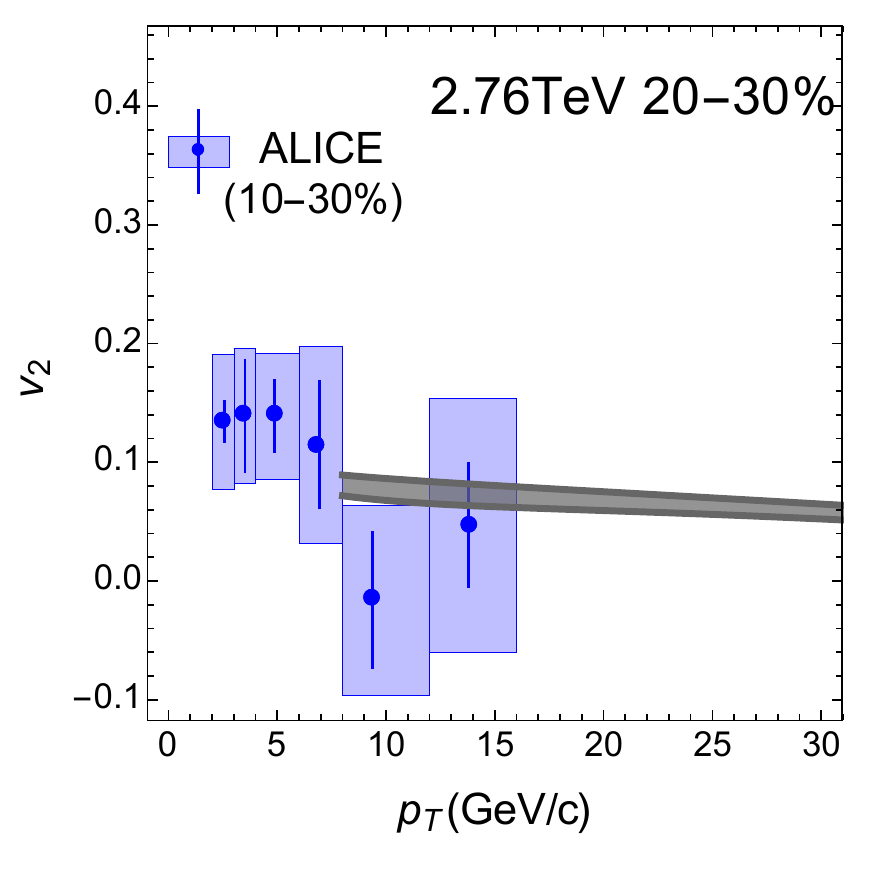} \hspace{0.02in} \\
\includegraphics[width=0.35\textwidth]{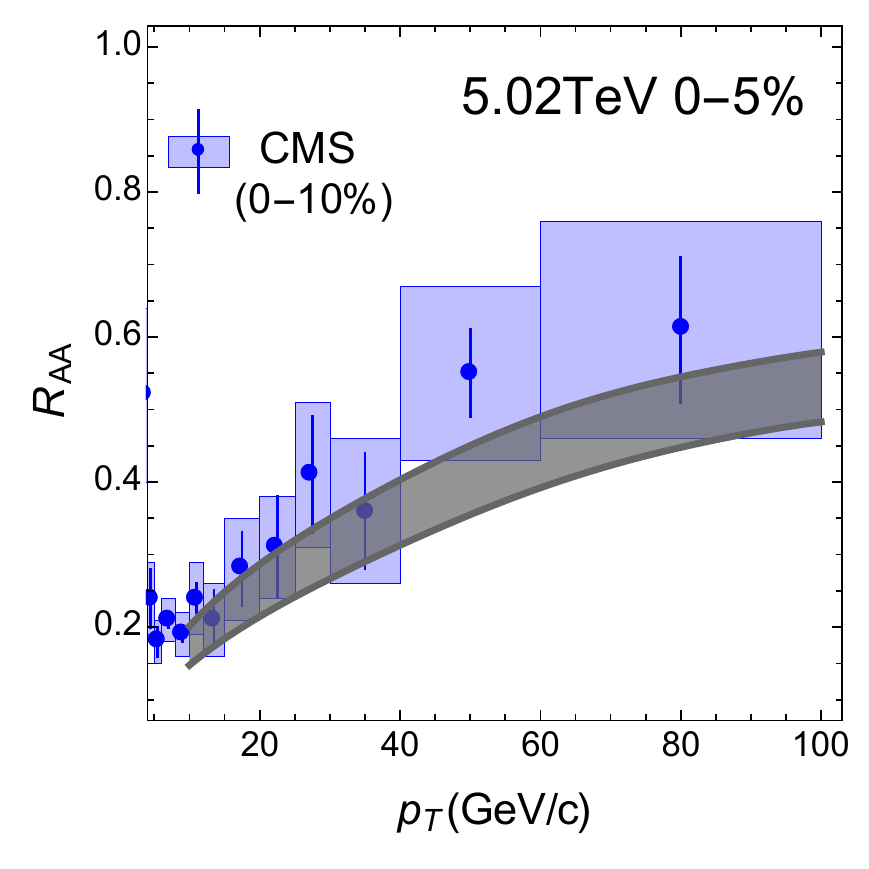} \hspace{0.02in}
\includegraphics[width=0.35\textwidth]{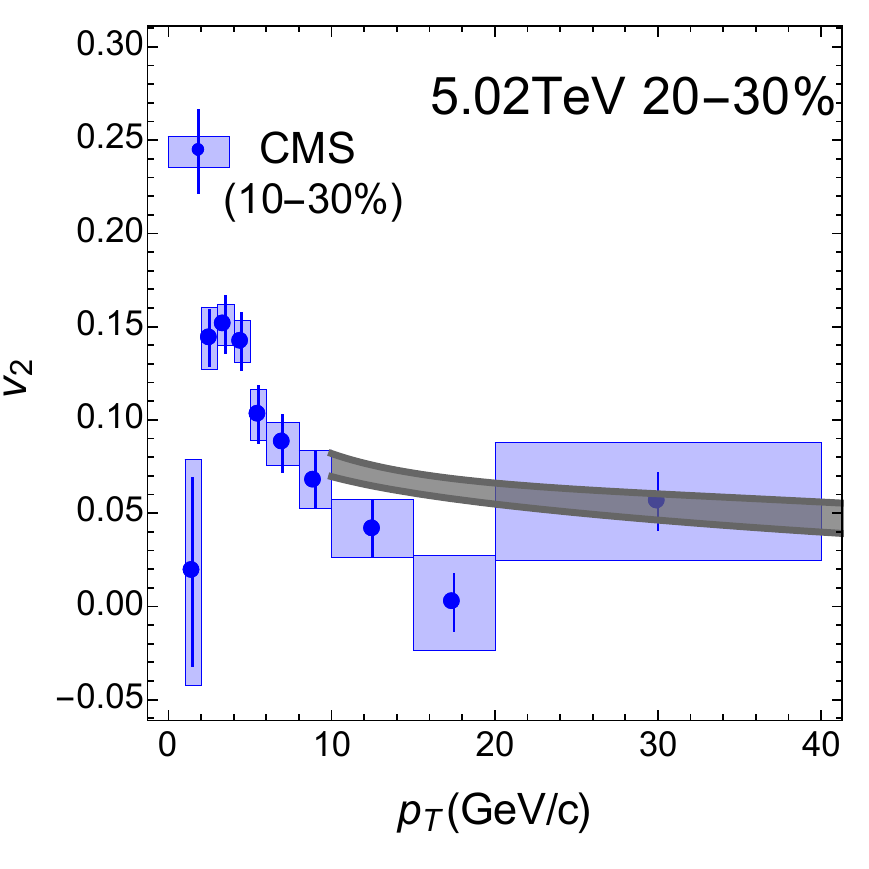}
\vspace{-0.2in}
\caption{(color online) Predictions from CUJET3 for central and semi-central $R_{AA}$ and $v_2$ of high $p_T$ D mesons at LHC 2.76TeV and 5.02TeV collisions, compared with currently available experimental data. }  \label{fig_heavy}
\end{center}
\vspace{-0.4in}
\end{figure}

In the above we've used only light flavor data to constrain the optimal values of key parameters in CUJET3. Therefore, the heavy flavor energy loss observables (e.g. high $p_T$ D meson $R_{AA}$ and $v_2$) can be used to further provide a critical  independent test of the CUJET3 framework. In Fig.~\ref{fig_heavy}  we compare CUJET3 results with available data for central and semi-central $R_{AA}$ and $v_2$ of high $p_T$ D mesons at LHC 2.76TeV and 5.02TeV collisions. One again observes very good agreement between model and data, validating a successful unified description of CUJET3 for both light and heavy flavor jet energy loss observables. With more accumulated statistics  and shrinking error bars as well as data from other collaborations, we expect more stringent future test from the heavy flavor sector to help further constrain CUJET3. 
\vspace{-0.2in}

\section{Summary}
\label{}

In summary, we've preformed a systematic test of the CUJET3 jet energy loss modeling framework with global high $p_T$ hadron observables in heavy ion collisions. The CUJET3, based on a nonperturbative microscopic picture of the medium as semi-quark-gluon-monopole plasma, has succeeded in providing a united description for comprehensive sets of data, from average suppression to azimuthal anisotropy, from light to heavy flavors, from RHIC 200GeV to LHC 2.76TeV as well as 5.02TeV collisions. We envision nontrivial further exploration of event-by-event jet energy loss study~\cite{Noronha-Hostler:2016eow,Betz:2011tu,Zhang:2012ha,Zhang:2012ie} via CUJET3 framework in the near future which shall allow quantifying higher harmonic anisotropy coefficients of high $p_T$ hadrons.

{\bf Acknowledgments.} 
The research of SS, JX and JL  is supported in part by  the NSF Grant No. PHY-1352368. MG and JL also acknowledge partial support by the IOPP, CCNU, Wuhan, China. 
\vspace{-0.2in}




\bibliographystyle{elsarticle-num}
\bibliography{<your-bib-database>}



\end{document}